# Photoconductivity effects in mixed-phase BSCCO whiskers

M Truccato[1,*], D Imbraguglio[2], A Agostino[3], S Cagliero[3], A Pagliero[1], H Motzkau[4], A Rydh[4]

1. NIS Centre of Excellence, Dipartimento di Fisica, and CNISM-UdR Torino Università, Via Pietro Giuria 1, I-10125, Torino, Italy

2. INRIM-Istituto Nazionale di Ricerca Metrologica, Strada delle Cacce 91, I-10135 Torino, Italy

3. NIS Centre of Excellence, Dipartimento Chimica Generale e Chimica Organica, and CNISM UdR Torino Università, C.so Massimo D'Azeglio 48, I-10125, Torino, Italy

4. Department of Physics, Stockholm University, AlbaNova University Center, SE-10691 Stockholm, Sweden

## Abstract

We report on combined photoconductivity and annealing experiments in whisker-like crystals of the Bi-Sr-Ca-Cu-O (BSCCO) high-$T_c$ superconductor. Both single-phase $Bi_2Sr_2CaCu_2O_{8+\delta}$ (Bi-2212) samples and crystals of the mixed phases $Bi_2Sr_2Ca_2Cu_3O_{10+x}$ (Bi-2223)/Bi-2212 have been subjected to annealing treatments at 90°C in air in a few hours steps, up to a maximum total annealing time of 47 h. At every step, samples have been characterized by means of electrical resistance vs temperature ($R$ vs $T$) and resistance vs time at fixed temperature ($R$ vs $t$) measurements, both in the dark and under illumination with a UV-VIS halogen arc lamp. A careful comparison of the results from the two techniques has shown that, while for single-phase samples no effect is recorded, for mixed-phase samples an enhancement in the conductivity that increases with increasing the annealing time is induced by the light at the nominal temperature $T$ = 100 K, i.e. at an intermediate temperature between the critical temperatures of the two phases. A simple pseudo-1D model based on the Kudinov's scheme [Kudinov *et al.*, Phys. Rev. B **47,** 9017-28, (1993)] has been developed to account for the observed effects, which is based on the existence of Bi-2223 filaments embedded in the Bi-2212 matrix and on the presence of electronically active defects at their interfaces. This model reproduces fairly well the photoconductive experimental results and shows that the length of the Bi-2223 filaments decreases and the number of defects increases with increasing the annealing time.

* Corresponding author

1. Introduction

The interaction between electromagnetic radiation and transport properties of materials is an interesting and extensively studied topic. In semiconductors, for instance, an enhancement of the electrical conductivity is commonly observed as a consequence of illumination, which is also known as photoconductivity. Concerning superconductors, it was shown in the past that illumination with visible light induces detrimental effects on the superconductivity of materials with low critical temperature ($T_c$) due to the breakdown of the Cooper pairs, which results in a quasiparticle excess and therefore in an increased effective temperature of the system [1]. A similar behaviour has been detected also in high $T_c$ superconductors (HTSC), but only at sub-nanosecond time scales and very low photon doses [2]. On the contrary, when the excitation time scale is at least of the order of milliseconds and at higher photon doses ($\geq 10^{19}$ photon/cm$^2$), HTSC can exhibit an enhancement in their superconducting properties (e.g. $T_c$ and the critical current density, $J_c$) that persists after the light is turned off, provided that these materials are kept at low enough temperatures. Such a phenomenon is known as persistent photoinduced superconductivity (PPS) [3]. Moreover, it is also well-known that in these compounds the normal state can increase its electrical conductivity by many orders of magnitude upon visible illumination and that this increase has a very long, temperature dependent relaxation time [4]. For this reason, this phenomenon has been defined as persistent photoconductivity (PPC).

The HTSC family that has been most extensively studied from the point of view of PPS and PPC is the $RBa_2Cu_3O_x$ class (R = rare earth). Gilabert *et al.* [5] report a detailed review of the corresponding experimental features. A widely accepted, semiconductor-like interpretation for such effects has been proposed by Kudinov *et al.* [4]. They suggested that the electrons produced in the $CuO_2$ planes during the electron-hole pair photoexcitation process are trapped in the unoccupied *p* levels of O$^-$ ions localized in adjacent $CuO_x$ chains. Therefore, mobile holes are created in the extended states of the $CuO_2$ planes, increasing the electrical conductivity of the normal state and possibly leading to superconductivity if the carrier concentration is high enough. However, subsequent experiments have detected PPC also in $Tl_2Ba_2CuO_{6+\delta}$ [6], which has no $CuO_x$ chain in its crystal structure, showing that the Kudinov's model has to be extended in such a way to include other kinds of defects as possible electron traps.

Concerning the $Bi_2Sr_2Ca_{n-1}Cu_nO_{2n+4+x}$ (BSCCO, *n*=1, 2 or 3) system, there have been some reports about the existence of a small PPC for the phase with *n*=2 in presence of Y-doping (i.e. $Bi_2Sr_2Ca_{1-x}Y_xCu_2O_{8+\delta}$) [5,7].

Moreover, a photodoping effect has been observed in bicrystal grain boundary junctions (GBJ) of $Bi_2Sr_2CaCu_2O_{8+\delta}$ (Bi-2212), with corresponding changes in $T_c$, in the energy gap and in the normal state resistivity [8]. In the latter case, it has been considered that oxygen-deficient grain boundaries can very likely play the same role in Bi-2212 as the $CuO_x$ chains do in YBCO, acting as traps for the photoexcited electrons. However, neither PPS nor PPC was observed in homogeneous Bi-2212 systems such as thin films [5,9] or whiskers [10], which can be interpreted as a further confirmation of the validity of the Kudinov's model.

In this framework, it seems that a necessary condition for the possible observation of PPS or PPC in Bi-2212 is the presence of suitable defects representing effective electron traps. It is well-know that even in high quality Bi-2212 crystals such as whiskers, some defects like twinning, dislocations, growth steps or chemical impurities can be revealed by SEM and TEM analysis, especially in large samples [11]. Apparently, they are not effective from the point of view of PPC, however.

Another kind of defects that is quite common in Bi-2212 whiskers is the intergrowth of the phase corresponding to $n=3$ (also known as Bi-2223), as can be noticed from the existence of two or more transition steps in the resistivity versus temperature curves [12]. In most cases, this is considered an undesired feature of the material, which is probably the reason why only a few papers can be found in literature focusing on the understanding of multi-phase BSCCO sample properties. On the other hand, some of us have already shown that annealing a double-phase whisker (i.e. containing both Bi-2212 and Bi-2223) at room temperature in helium atmosphere induces a phase transformation in the material, changing the relative amount of the two phases and modifying their superconducting properties [13]. This is quite different from what happens in single-phase Bi-2212 whiskers, where an oxygen depletion process takes place [14].

It is reasonable to expect that multi-phase samples, being composed of different crystalline domains or intergrowths, contain many more defective grain boundaries than single-phase whiskers. This fact, along with the possibility of modulating the amount of the intergrowth regions by means of the annealing process, makes these samples good candidates for the observation of photoconductivity phenomena in BSCCO crystals, resembling the experimental conditions of the Bi-2212 GBJ's. However, no combined annealing and photoconductivity investigation in multi-phase BSCCO whiskers has been carried out so far, to the best of our knowledge. This is the purpose of the present experiment, where we have measured highly defective multi-phase samples starting from their as-grown conditions and studying their evolution during a mild temperature annealing process.

The outcomes of such measurements are expected to provide useful information to the field of visible and infrared radiation detection by means of HTSC nanowires, especially for the approach aiming at achieving photon number resolution via multiple nanowire elements connected in parallel. Indeed, in this case a careful design of the resistors in series with each nanowire is required to obtain a voltage signal proportional to the photon number, so that the possible change in the nanowire resistance induced by the radiation itself should also be taken into account [15].

Moreover, it should also be considered that, from a general point of view, the possibility of controlling the material photo-response lies at the core of optoelectronic devices. Therefore, the investigation of the modifications of HTSC behaviour induced by intergrowths and by thermal treatments represents a valuable tool to obtain further insight into the nature of photoactive defects and into their creation mechanism, which is an essential step towards the goal of their engineering.

2. **Experimental methods**

BSCCO whiskers were grown at 862-868°C following a procedure that is explained in detail in a previous paper [16]. The crystals were then mechanically removed from the glassy plates where they grew and selected under an optical microscope.

Crystals intended for photoconductivity measurements were placed onto a sapphire substrate with their *c*-axis perpendicular to the substrate plane. A stainless steel mask was placed on each crystal for the evaporation process, which resulted in a 2.5 μm thick silver layer plus a gold cap layer of 30 nm with a four-probe configuration. Electrical contact to the material was achieved after annealing the samples at 450°C for 5 minutes in $O_2$ gas flow [17]. Crystals dedicated to in-plane resistivity mapping were cleaved by gluing them between sapphire substrates by means of Stycast® 2651 epoxy resin with the *c*-axis of the whiskers perpendicular to the sapphire surfaces. About 20 nm of Ag and 200 nm of Au were e-gun evaporated immediately after the cleaving. Standard photolithography and low-current Ar ion milling were then carried out to obtain a six-probe geometry. Preliminary resistance versus temperature (*R* vs *T*) measurements were performed in a liquid nitrogen or in a home-built cryogen-free cryostat, while geometrical characterizations were obtained by means of SEM and AFM analyses. On the whole, seven samples were successfully produced and characterized by using these procedures.

The photoconductivity experiments were carried out by placing the samples at the center of a circular illuminated area (diameter = 40 mm) on the top of a copper cold head with cylindrical shape that was connected to a Leybold RW5 compressor. The temperature was controlled by means of a Leybold LTC 60 temperature controller and monitored by two silicon diodes placed at different positions. One of them ($T_b$) was located as near the crystal as possible ($\approx$ 8 mm), while the other ($T_a$) was placed at the border of the light beam and represented the nominal temperature of crystal, as discussed in the following. Sample illumination was obtained by means of a 70 W discharge halogen arc mercury lamp able to deliver a power density of about 0.4 W cm$^{-2}$ at the sample position. The lamp spectrum ranges from 230 to 770 nm and the quartz optical window of the cryostat has no absorption line in this range. $R$ vs $T$ data were collected on this system both in dark and in light conditions over the temperature range $T$ = 65-295 K. All these measurements were performed on warming with a typical temperature ramp rate of 1.2 K min$^{-1}$ to minimize thermal hysteresis. No care was taken in shielding the samples from the earth's magnetic field. Complementary information was obtained by stabilizing the sample temperature at different values and by turning the illumination on and off while monitoring the material resistance as a function of time ($R$ vs $t$ characterization). These measurements were acquired starting with recording the resistance values in dark conditions for a period of 5 minutes in order to check for possible thermal drift phenomena, then the lamp was turned on for about 25 minutes and finally turned off for 20 minutes. This procedure was repeated at least for the sample nominal temperatures $T_a$= 70, 100 and 110 K.

In order to analyze the evolution of the photoconductivity phenomena, $R$ vs $T$ and $R$ vs $t$ measurements were taken on the samples both in their as-grown state and after each step of the ageing process. Whiskers were aged by extracting them from the cryocooler and undergoing a thermal annealing at 90°C in air, in steps of variable durations according to the ageing stage of each crystal.

Measurements intended for in-plane resistivity mapping were performed on a 17 T cryogen-free system by Cryogenic Ltd equipped with a 8-channel PXI multi-component acquisition system including FPGA and low-noise preamplifiers. A lock-in measurement technique has been used achieving a noise level of a few nanovolts. *R* vs *T* data acquisitions were simultaneously carried out in the different portions of the crystals in the temperature range between about 40 K and 115 K.

### 3. Results

#### 3.1 Dark measurements

Preliminary characterizations showed that three out of the seven samples consisted of pure Bi-2212 phase, while the four remaining contained both Bi-2212 and Bi-2223. Many of the contacts broke down at the initial stages of the photoconductivity experiment due to the cryogenic thermal cycles and to the ageing steps undergone by the samples. In the following, we therefore focus on the sample whose electrical contacts survived the longest series of experimental steps, providing the most significant results.

**Table 1**. Details of the annealing treatments performed on the most significant sample (WBSCG_4). Each step is described in terms of total annealing time $t_{ann}$, annealing temperature $T_{ann}$ and of the corresponding critical temperatures.

| Ageing step | $t_{ann}$ (h) | $T_{ann}$ (°C) | $T_{c1}$ (K) | $T_{c2}$ (K) | $T_{c3}$ (K) |
|---|---|---|---|---|---|
| a | 0 | 90 | 76.9 | 104.4 | 93.8 |
| b | 1 | 90 | 77.8 | 104.7 | 95.1 |
| c | 2 | 90 | 78.6 | 104.7 | |
| d | 17 | 90 | 78.5 | 104.5 | |
| e | 32 | 90 | 79.0 | 104.5 | |
| f | 47 | 90 | 79.6 | 103.9 | |
| g | 1 year | 25 | 79.1 | 106.6 | 92.5 |

Table 1 reports the list of the thermal treatments this sample underwent. After the first set of two 1 h long annealing steps, a further set of treatments was carried out increasing the dwell time for each step up to 15 h, in order to accelerate the process and to obtain information on the evolution of the material on an extended timescale. Five annealing steps at 90°C were completed and measured, corresponding to 47 h of total annealing time. In addition, one more measurement was taken in dark conditions after storing the sample for one year at room temperature, which is indicated in Table 1 as the last annealing step.

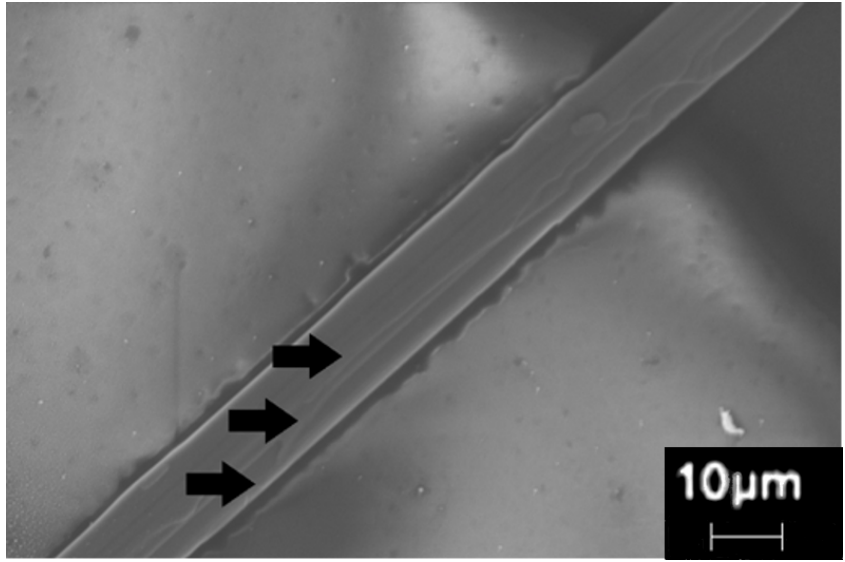

**Figure 1.** SEM micrograph of sample WBSCG_4 acquired at the end of the whole cycle of thermal treatments. The black arrows indicate three different growth steps on the whisker surface.

Figure 1 shows the sample at the end of the whole ageing process. The picture corresponds to the portion of the whisker between the two voltage contacts, which is the one sensed by the resistance measurements. It is possible to observe that at least three distinct growth steps exist on the crystal top surface, as highlighted by the black arrows. Size measurements resulted in a crystal cross section $S = 35.5 \pm 0.8$ µm$^2$, while the distance between the two voltage contacts mid-points is $L = 147.9 \pm 1.2$ µm. These values were used as the proper geometrical factors for the conversion of the $R$ measurements into values of the effective resistivity of the material $\rho_{eff}$, defined as $\rho_{eff} = R\,S\,/L$. Because of the insensitivity of the BSCCO whisker geometry to the entanglement of the in-plane ($\rho_{ab}$) and out-of-plane ($\rho_c$) electrical resistivity components [18,19], the $\rho_{eff}$ values can be considered as representative of the in-plane conduction process in the whiskers.

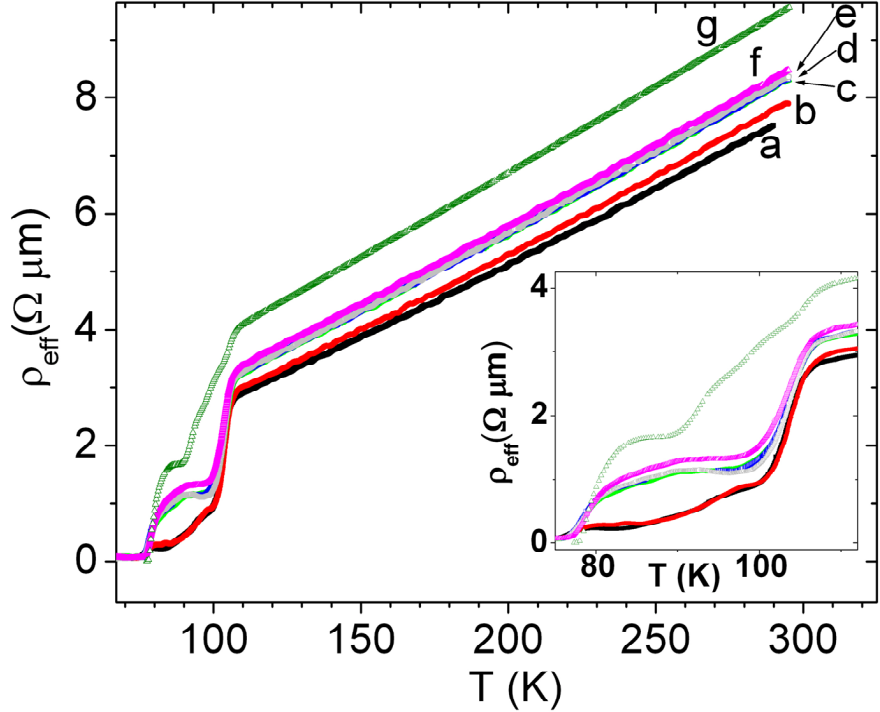

**Figure 2**. (Color online) $\rho_{eff}$ vs $T$ characteristics in dark conditions. Curve labels correspond to annealing steps listed in Table 1. The inset shows a blow-up of the transition region.

The electrical characteristics measured in dark conditions during the whole annealing experiment are shown in Fig. 2. The inset represents a blow-up of the transition region. Critical temperatures have been defined as the inflection points of the curves and have been determined by means of their numerical derivative. The corresponding values are listed in Table 1 and plotted in Fig. 3.

It is apparent that multiple transitions occur, revealing that the material consists of at least two phases. The first transition temperature, labeled as $T_{c1}$, ranges from 76.9 K to 79.6 K and can be identified with the typical critical temperature of the Bi-2212 phase resulting from the whisker production method [16,17,20]. The second transition temperature, indicated as $T_{c2}$, ranges between 103.9 and 106.6 K, and therefore corresponds to typical values of the Bi-2223 phase in mixed-phase whiskers [13]. Apparently, the amount of this phase does not exceed the percolation threshold, since the corresponding transition is not complete. Finally, a careful analysis of the curve derivatives reveals that at some stages of the ageing process (i.e. at steps *a*, *b* and *g*) a third transition temperature, labeled as $T_{c3}$, can be detected, with values between 92.5 and

95.1 K that correspond to the Bi-2212 phase with both optimal doping and optimal cationic ratio [21]. It has already been shown in previous ageing experiments on whiskers that such values can be ascribed to the presence of Bi-2212 phase domains resulting from the decomposition of the Bi-2223 phase [13]. This possible interpretation will be further discussed later on. For the moment, we just want to stress that systematic investigation by XRD of the relationship between average crystal structure and whisker growth conditions has shown that in these systems the Bi-2223 phase volume fraction is always below the detection threshold (i.e. less than about 5%) [16]. Moreover, local mapping of the whisker crystal structure by synchrotron radiation microprobe has also shown that the presence of more than a single Bi-2212 domain in such whiskers is typically correlated with anomalous profiles in the AFM maps [14], which have not been detected in the present case.

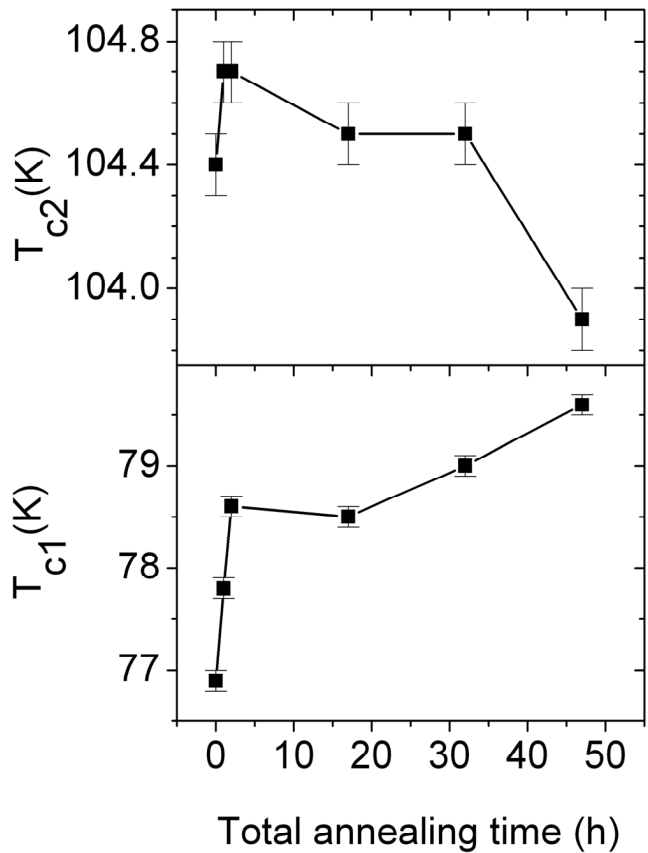

**Figure 3**. Evolution of the critical temperatures $T_{c1}$ and $T_{c2}$ with the annealing time. Error bars represent the uncertainties in the determination of the inflection points.

The comparison between the curves of the different ageing steps reveals a monotonic increase of the $\rho_{eff}$ values with increasing the annealing time, which is somewhat irregular in the temperature range below $T_{c2}$ and steadier for higher temperatures. Concerning the critical temperatures, Fig. 3 shows that for $T_{c2}$ only a very weak variation can be detected, while $T_{c1}$ definitely increases with increasing the annealing time. These behaviours reveal that some process involving minor changes in the doping states of both the Bi-2223 and the Bi-2212 phases occurred during the experiment. A similar situation has already been observed in a previous ageing experiment of a mixed Bi-2212 and Bi-2223 phase whisker [13], where resistivity data could be well explained by a process consisting in the local disruption of the Bi-2223 phase that released O ions available for a change in the doping level of the Bi-2212 phase.

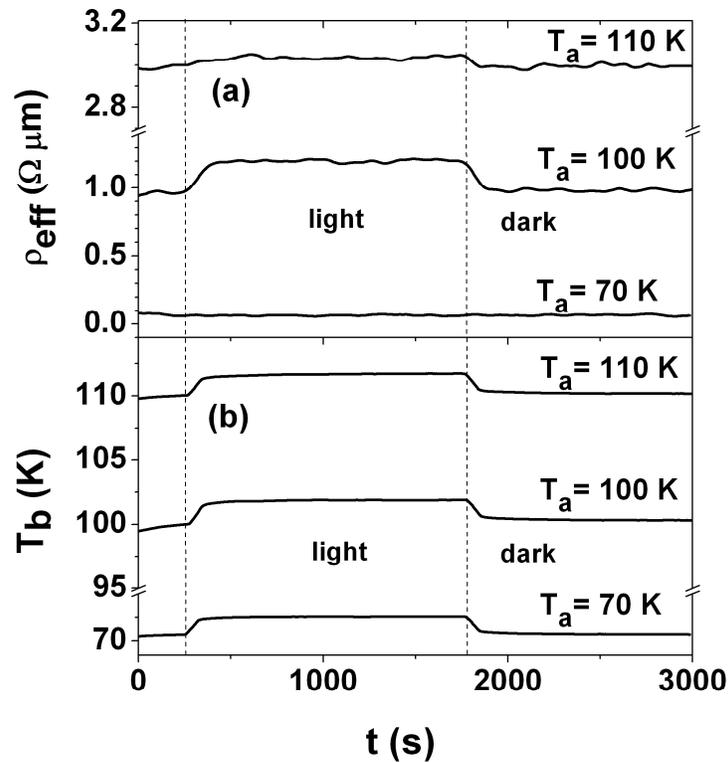

**Figure 4**. Time evolution in as-grown conditions (i.e. ageing step *a* of Table 1) of: (a) the sample effective resistivity $\rho_{eff}$, and (b) the temperature $T_b$ at a distance of about 8 mm from the sample (i.e. inside the illuminated area) for the nominal temperatures $T_a$ = 70 K, 100 K and 110 K. Vertical lines indicate the instants when the lamp was turned on or off.

### 3.2 Time behaviours and temperature profile

Both the electrical resistance of the sample and the temperature distribution around it have been monitored as a function of time and illumination at the nominal temperatures $T_a$ = 70 K, 100 K and 110 K, and this investigation has been performed at every stage of the ageing process. Figure 4 shows a typical example of such measurements while switching the illumination on and off. The behaviour of $T_a$ has not been reported because it is constant in time within our resolution limit of 0.02 K and therefore insensitive to the presence or absence of light (i.e. $T_{a,l} = T_{a,d}$), which makes $T_a$ suitable to represent the nominal temperature of the system. On the other hand, Fig. 4(b) points out that, at a distance of 8 mm from the sample, illumination induces local heating that disappears when the lamp is turned off. Typical temperature increase for $T_b$ under illumination is $T_{b,l} - T_{b,d}$ = 1.87 K, 1.57 K and 1.53 K for $T_a$ = 70 K, 100 K and 110 K, respectively, and takes place with characteristic times ranging from 60 s to 67 s, depending on the nominal temperature. It is worth noticing that both the temperature variations and the time constants of the heating process are the same also during the cooling stage occurring when the lamp is switched off, confirming that the light induces a completely reversible heating in the system.

The time evolution of the effective resistivity of the sample $\rho_{eff}$ is presented in Fig. 4(a). It is possible to observe that at $T_a$ = 70 K the material preserves its vanishing resistivity both in dark conditions and under illumination, showing that at the sample position the temperature increase induced by the light (i.e. $T_{s,l} - T_{s,d}$) is not large enough to drive it out of the superconducting state. On the contrary, at $T_a$ = 100 K and $T_a$ = 110 K a resistivity increase $\rho_{eff,l}(T_a) - \rho_{eff,d}(T_a)$ is observed under illumination, corresponding to a relative amount of 23% and 2%, respectively, for the plateau values. Although this is expected from a qualitative point of view because of the heating induced by the light, a quantitative discussion of the phenomenon requires further efforts.

In this sense, an important issue is represented by the knowledge of the exact temperature at the sample position when light illuminates the experimental setup $T_{s,l}$. Indeed, some temperature gradient is induced by the thermal load of the light, so that in steady state conditions the temperature $T_{s,l}$ of the sample, which is positioned at the center of the illuminated area, is expected to be higher than the temperature $T_{b,l}$, which is measured at the closest point compatible with the experimental limits (i.e. 8 mm away from the crystal position). A reasonable model for the estimation of $T_{s,l}$ that approximates the real conditions of the illuminated area consists in the assumption of cylindrical symmetry around the optical axis of the system. Under this assumption and in steady state conditions, the Fourier equation for heat conduction and the conservation of energy applied to the copper head lead to the parabolic temperature profile (see Appendix):

$$T(r) = -\frac{P}{4Kl}r^2 + C \quad , \qquad (1)$$

where $P$ is the power per unit surface delivered by the light beam, $K$ is the copper thermal conductivity, $l$ is the copper head thickness, $r$ is the distance from the beam center, and $C$ is an integration constant that can be determined by applying a proper boundary condition. To this purpose, the knowledge of $T_{b,l} = T(r=8~\text{mm})$ under illumination can be exploited, along with the useful observations that in the dark $T_{b,d} \approx T_a$ due to thermal equilibrium and that $T_a$ is constant irrespectively of the illumination conditions. By inserting this information and the values of the experimental parameters ($P = 400$ W m$^{-2}$, $K = 615$-$449$ W K$^{-1}$ m$^{-1}$, $l = 2$ cm) into Eq. (1), the sample real temperature $T_{s,l}$ can be estimated as $T(r = 0)$. The results show that $T_{b,l}$ underestimates $T_{s,l}$ by about 1 K: more specifically $T_{s,l} - T_{b,l} = 0.98$–$1.24$ K, depending on the experimental run. The reliability of this estimation procedure is further testified by the fact that the maximum variation of $T_{s,l}$ under the same experimental conditions (i.e. at fixed $T_a$) is equal to 0.67 K, which is fully comparable to the corresponding variability of 0.43 K experimentally recorded for $T_{b,l}$.

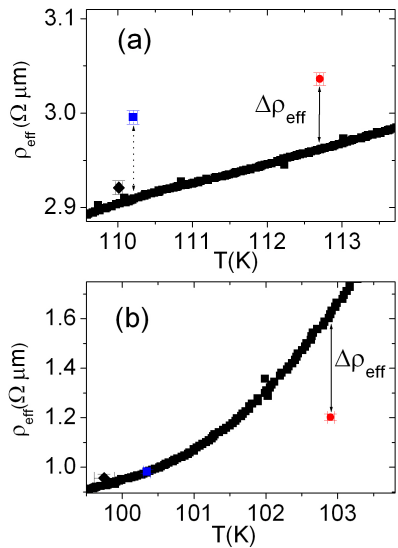

**Figure 5**. (Color online) Comparison between the dark measurements of $\rho_{\mathit{eff,d}}$ vs $T$ of ageing step *a* (black solid squares, same data of Fig. 2) and the corresponding values of $\rho_{\mathit{eff}}$ vs $t$ for: (a) $T_a = 110$ K, and (b) $T_a = 100$ K. Black solid diamonds represent the average sample conditions of Fig. 4 *before* turning the light on, red solid circles describe the situation *during* illumination and blue solid squares *after* turning the light off.

### 3.3 Photoconductivity behaviour

The knowledge of the sample temperature both in the dark and under light conditions allows obtaining more information about the influence of the light on the conduction process. The time-averaged values of $\rho_{\mathit{eff}}$ and $T_b$ *before* and *after* illumination can be deduced from the data of Fig. 4(a) to describe the sample situation in both conditions as $\rho_{\mathit{eff},d}$ ($T_{s,d}$), while the same averages for $\rho_{\mathit{eff}}$ and $T_{s,l}$ taken during the illumination period represent the status of the sample under the light by means of the $\rho_{\mathit{eff},l}(T_{s,l})$ data points. Such values have to be compared to the $\rho_{\mathit{eff},d}$ ($T$) curves acquired in dark conditions, as shown in Fig. 5 in the case of ageing step *a*. It can be noticed that there is good agreement between the values of $\rho_{\mathit{eff},d}$ monitored in time *before* illumination and the $\rho_{\mathit{eff},d}$ vs $T$ characteristics, testifying that both measurements are describing the same material conditions. On the other hand, data taken under illumination are non-negligibly different from the ones in dark conditions, resulting both in higher and lower values than expected for $T_a$ = 110 K and $T_a$ = 100 K, respectively. Finally, it is possible to observe that, after switching the light off, the material fully recovers the dark condition status in the case of nominal temperature $T_a$ = 100 K, while for $T_a$ = 110 K remarkably higher values than the ones in the dark are retained. However, this behavior is not typical and limited to $T_a$ = 110 K for ageing steps *a* and *b*. It is also worth mentioning that the behaviour shown in Fig. 5(b) was not restricted to this sample only, but was observed in other cases as well.

For a quantitative evaluation, under light conditions we can define the photoinduced variation of the sample effective resistivity $\Delta\rho_{\mathit{eff}}$ as:

$$\Delta\rho_{\mathit{eff}}\left(T_{s,l}\right) = \rho_{\mathit{eff},l}\left(T_{s,l}\right) - \rho_{\mathit{eff},d}\left(T_{s,l}\right) \quad . \quad (2)$$

Such a definition can be applied to all the experimental conditions of the study, i.e. to the different values of nominal temperature and to the subsequent stages of the annealing process. The corresponding results are shown in Fig. 6(a) as a function of the total annealing time for the sample temperatures $T_{s,l}$ = 103.0±0.3 K and 112.8±0.2 K, corresponding to $T_a$ = 100 K and 110 K, respectively. A definition analogous to Eq. (2) can also be applied to investigate the persistency of photoconductivity after illumination, just with $\rho_{\mathit{eff},l}$ ($T_{s,l}$) substituted by the average of the $\rho_{\mathit{eff}}$ values measured in the dark from 200 to 1200 s *after* turning the lamp off. The corresponding results are shown in Fig. 6(b).

Measurements corresponding to $T_a = 70$ K ($T_{s,l} = 73.8.0\pm0.3$ K) have not been reported for clarity, being always indistinguishable from zero both under and after illumination. This is in agreement with the fact that in this case $T_{s,l}$ is less than $T_{cl}$ and therefore the sample remains in its superconducting state both in dark and in light conditions. On the other hand, data for $T_a = 100$ K under illumination show that a decrease in resistivity takes place, starting with non vanishing values for the pristine sample and then increasing the size of the effect in a way that is roughly proportional to the annealing time. The percentage of resistivity decrease ranges from 41% to 57% of the value in dark conditions, depending on the ageing step. This is a novel experimental result for BSCCO. A further evidence of the reliability of this result is represented by the fact that, even under the extreme (and unlikely) assumption that $T_{s,l} = T_{b,l}$, the estimated values of $\Delta\rho_{eff}$ are reduced by a factor ranging from 2 to 6 (depending on the ageing step) but do not vanish. Moreover, Fig. 6(a) shows that at $T_a = 110$ K no significant effect exists under illumination a part from ageing steps *a* and *b*.

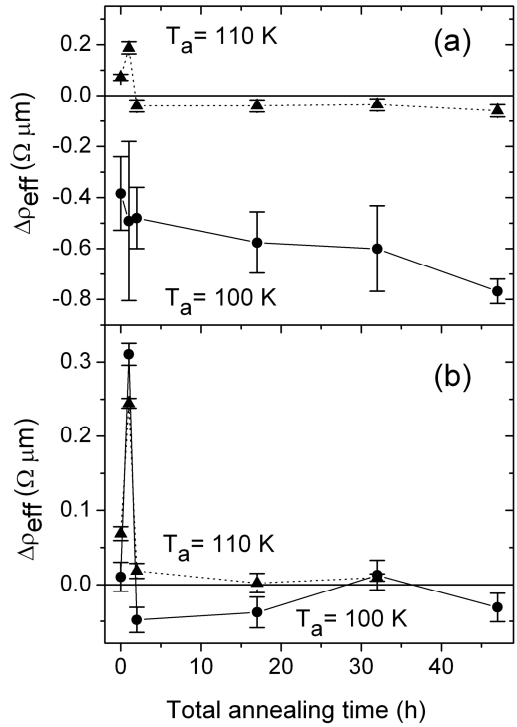

**Figure 6**. Photoinduced variation of the sample effective resistivity $\Delta\rho_{eff}$ as a function of the annealing time: (a) *under* illumination, and (b) *after* illumination. Solid triangles and circles correspond to the nominal temperature $T_a = 110$ K and 100 K, respectively. Error bars have been determined by propagating the uncertainties in the estimation of $T_{s,l}$.

Concerning the persistency of photoconductivity, Fig. 6(b) shows that over the investigated timescale a non-vanishing effect can be detected only for the annealing steps *a* and *b*, which are the same where some photoconductivity effect can be observed at $T_a$ = 110 K. Combining such general trend with the observation from Fig. 4 that the decay time constants for $T_b$ and $\rho_{eff}$ look very similar, it can be concluded that the photoconductivity characteristic decay time should be much less than the thermal time constant of our experiments.

We have also used the same procedure to determine $\Delta\rho_{eff}$ under illumination for the samples consisting of pure Bi-2212 phase. In this case it turns out that $\Delta\rho_{eff}$ = 0 within uncertainties for all the ageing steps experimentally available, whatever temperature is investigated, which confirms the results of [10].

## 4. Discussion and modelization

The experimental evidences reported above clearly indicate that the photoconductivity effect is closely related to the presence of two phases in BSCCO crystals and that can be promoted by the annealing process. Moreover, the promotion of photoconductivity by annealing is associated with changes in the electrical features of the Bi-2212 phase (e.g. $T_{c1}$).

It is worth noticing that similar changes have already been observed for Bi-2212 in mixed phase crystals, showing that a phase transformation can take place for Bi-2223 in favour of Bi-2212 [13,22]. This typically results in a general doping effect for the latter phase with the possible segregation of crystalline domains with slightly different doping properties.

Therefore, the presence of different domains induces to consider some mechanism of the Kudinov type as the possible explanation for the photoconductive effect, since it has been demonstrated that domain boundaries can include amorphous regions [23] and represent extended non-stoichiometric, oxygen defective areas that should have strong effect on the transport properties [24], possibly resulting in effective electron traps. Actually, this kind of mechanism has also been invoked by Gilabert *et al*. [8] to explain their data on Bi-2212 bicrystal grain boundary junctions and is supported by our observation that the photoconductivity effect is visible only when at least two phases are present in the crystals.

In order to formulate a possible model for data interpretation, the definition of the topology of the different domains and of their boundaries is a crucial issue. From this point of view, HRTEM observations have shown that Bi-2223 inclusions in a Bi-2212 matrix typically consist of 10-25 nm long filaments that are separated by 1 nm or less in the *ab*-plane direction and from 1 to 40 Bi-2212 unit cells along the *c*-axis direction [25,26]. The same authors have also demonstrated that, in the temperature range between $T_{c1}$ and $T_{c2}$, such a structure induces a coupling of the filaments by proximity effect in the *ab*-plane direction and by Josephson effect along the *c*-axis. This coupling can result in a Bi-2223 superconducting percolating path for low enough magnetic fields or feeding currents and is destroyed at higher field or current values.

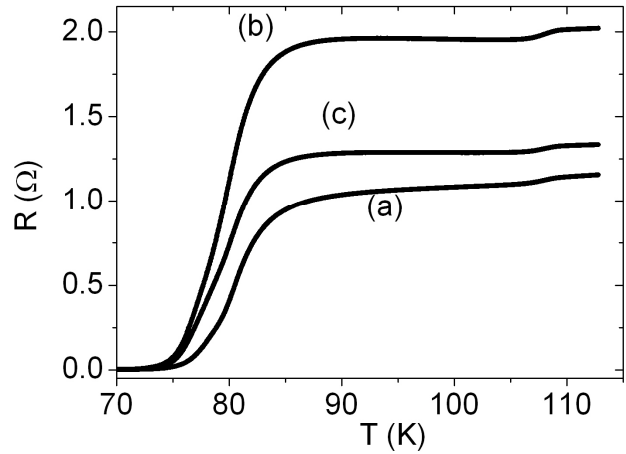

**Figure 7**. *R* vs *T* behaviour measured along the *a*-axis direction with $I = 50$ μA in a six-probe sample. Current was injected and extracted at the most external contacts. Curves (a), (b) and (c) correspond to the second, third and fourth longitudinal sections, respectively (i.e. the most central ones), out of the five into which the sample is divided by the electrical contacts.

We investigated the Bi-2223 topology in our samples by mapping their in-plane resistivity. Figure 7 displays the *R* vs *T* behaviour for three adjacent longitudinal sections of a six-probe sample prepared for this purpose. Sample width and thickness are 56.1±0.5 μm and 440±15 nm, respectively. The sample is divided by the contacts into five longitudinal sections, each of them 12±1 μm long, and the ones corresponding to curves (a), (b) and (c) of Fig. 7 are placed 11, 24 and 12 μm far away from the closest current contact, respectively. Such a geometry implies that these *R* vs *T* data basically represent the temperature behaviour of the in-plane resistivity $\rho_{ab}$, with only a minor contribution from the out-of-plane component $\rho_c$ [18]. The presence of a transition at $T_{c2}$ = 107.9 K in each curve testifies the fact that some amount of the Bi-2223 phase exists in each longitudinal section, but at the same time such amount is so small that the Bi-2223 material is not able to percolate through the Bi-2212 matrix for the whole section length. Therefore, our electrical measurements confirm the picture of Bi-2223 filamentary inclusions of [25], so that the spatial distribution of the Bi-2223 phase domains in our samples can be properly sketched like in Fig. 8(a).

Of course, each domain boundary represents a highly defective region that could take part in the photoconductivity process and the defect distribution is expected to be uniform along the boundary. However, electrical conduction in whiskers is essentially a 1D process and therefore we can approximate the real situation of the crystal by a pseudo-1D electrical model that groups together all the randomly distributed Bi-2223 filaments into an equivalent single filament with length $l_3$ and width $w_3$ (see Fig. 8(b)). It is also assumed for simplicity that every phase domain boundary extends across the whole sample thickness $t$. Concerning the defects associated with the domain boundaries, in a Kudinov-like scheme they should influence the conductivity properties by trapping the photoexcited electrons and therefore by acting as local sources of mobile holes that represent an additional amount of carriers. By assuming that each defect can trap one electron on the average, the number of additional carriers induced by the light is equal to the number of defects in the sample $N_{def}$. Such carriers are accelerated by the electric field in the crystal length direction. Obviously, both the inclusions and the matrix can be affected in their conduction properties only by those carriers that can reach them, which means that electrically active defects have to be located in the boundaries with a transverse alignment with respect to the electrical field. In our effective pseudo-1D electrical model, this means that the defects are distributed along both transverse boundaries of the Bi-2223 filament, in equal amounts for symmetry reasons (see Fig. 8(b)).

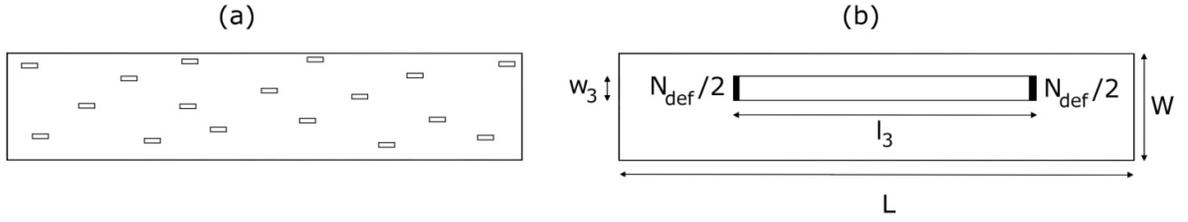

**Figure 8**. (a) Schematic representation of the real spatial distribution of the Bi-2223 filaments in a Bi-2212 matrix. (b) Equivalent electrical model obtained by grouping the filaments together in a single inclusion. $W$ is the crystal width, $L$ represents its length, $w_3$ and $l_3$ are the Bi-2223 phase filament width and length, respectively. The defects are grouped in two amounts of $N_{def}/2$ located at both transverse boundaries of the Bi-2223 filament.

Once the space distribution of the two phases has been established, the implications of the model can be evaluated for the different experimental situations. All of the photoconductivity measurements have been performed in the earth's magnetic field, i.e. $B_{exp} \approx 50$ μT. According to the data available for the lower critical field $B_{c1}$ in Bi-2212 [27] and Bi-2223 [28] and to the values of $T_{c1}$, $T_{c2}$, $T_{s,l}$, and $T_{s,d}$ reported above, such a magnetic field value implies that for $T_a = 70$ K both crystal domains are always in the Meissner state, whatever the illumination conditions. This means that no dissipation takes place anywhere in the sample and therefore no conductivity enhancement can be expected for $\rho_{eff}$ because of the photogenerated carriers, in agreement with our experimental observations.

The situation is different for $T_a = 100$ K. Let us indicate with $\rho_2$ and $\rho_3$ the resistivities of the Bi-2212 and Bi-2223 phase domains, respectively. In dark conditions, the fact that $T_{s,d} = T_a < T_{c2}$ and $B_{exp} < B_{c1}$ for Bi-2223 reveals that this domain should remain in the Meissner state, so that $\rho_3 = 0$. On the other hand, for the Bi-2212 phase $T_{s,d} > T_{c1}$, which implies that this phase takes its normal state resistivity value $\rho_{2,d} = \rho_2(T_{s,d} = 100$ K$)$. The corresponding spatial distribution of the resistivity is represented in Fig. 9(a).

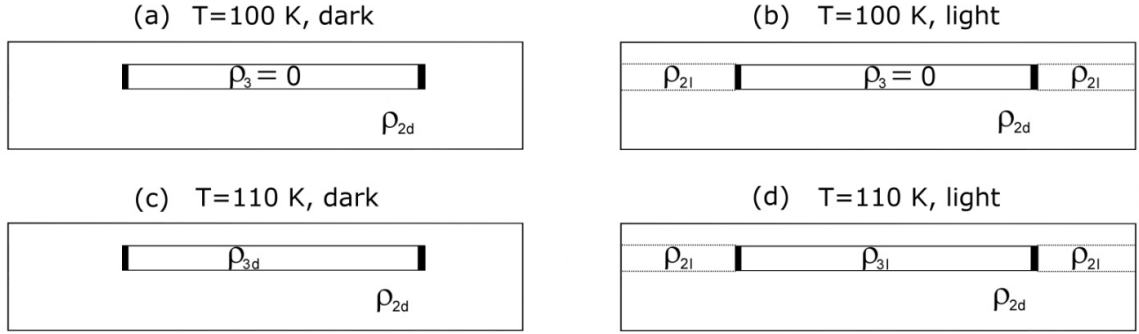

**Figure 9.** Spatial distribution of the different phase domains and of the regions affected by photogenerated carriers for: (a) $T_a = 100$ K in the dark, (b) $T_a = 100$ K under illumination, (c) $T_a = 110$ K in the dark and (d) $T_a = 110$ K under illumination.

When light is turned on at this temperature, changes consist in: i) the increase in the sample temperature due to the thermal load by the light, and ii) the photogeneration of mobile holes because of electrons trapped by the defects.

Concerning the temperature variation, its average value at $T_a = 100$ K is about 3.0 K. This amount is not enough to drive the Bi-2223 domain out of the Meissner state [28], so that $\rho_3 = 0$ still holds. On the other hand, strictly speaking, all the volume occupied by the Bi-2212 phase changes its resistivity value to $\rho_2(T_{s,l}) = \rho_2(T_{s,d} + 3$ K$)$. However, it has already been measured in pure Bi-2212 phase whiskers with different doping levels that the corresponding percentage increase of the in-plane resistivity $\rho_{ab}$ is 3.1% only [16], so that we can approximate $\rho_2(T_{s,l}) \approx \rho_2(T_{s,d}) = \rho_{2,d}$ at the leading order. This is equivalent to stating that in our electrical model the small temperature increase induced by the light can be safely neglected.

From the point of view of carrier photogeneration, some more hypotheses are necessary to quantify its effect. According to Kudinov's scheme, holes are created at the transverse phase boundaries of the sample. A basic assumption of our model is that photogenerated carriers can be accelerated by the electric field only, which is always oriented along the longitudinal direction in our experiment. This implies that no transport process is considered along the transverse direction, diffusion current included, making our model inherently 1D for the sake of simplicity. When an electric field is present, holes are injected in the downstream portions of the material and modify their conductivity properties by increasing the local carrier density. On the contrary, if no electric field is present in some region, the corresponding holes are not accelerated and cannot contribute to the conduction process as additional carriers. This consideration statistically implies that half of the photogenerated holes are ineffective because they originate from a Bi-2223 extremity that has no electric field downstream of it ($\rho_3 = 0$). However, even if they could take part to the conduction process, they would not be able to decrease the Bi-2223 phase resistivity since it already vanishes. For the remaining half, the only self-consistent choice is that they are injected in all of the crystal length that is not occupied by the Bi-2223 phase, i.e. $L - l_3$. Indeed, on one hand this choice best resembles the real situation of the crystal, where each extremity of the continuous, randomly distributed Bi-2223 micro-filaments is surrounded by a large portion of the Bi-2212 phase matrix where holes can be injected. On the other hand, such assumption summarizes the fact that our model is independent of the position of the single Bi-2223 filament in the Bi-2212 matrix, which can be mathematically proved. Therefore, all of the additional $N_{def}/2$ holes are considered to be spread over a volume that is $w_3$ wide and $L - l_3$ long (with a uniform number density for simplicity), resulting in a new resistivity value $\rho_{2,l}$ for this portion of the Bi-2212 phase. The corresponding situation is represented in Fig. 9(b).

At $T_a = 110$ K in the dark, both the Bi-2212 and the Bi-2223 phase are in their normal state because $T_{s,d} = T_a > T_{c1}$ and $T_{s,d} = T_a > T_{c2}$. Therefore their resistivity values are $\rho_{2,d} = \rho_2(T_{s,d} = 110$ K) and $\rho_{3,d} = \rho_3(T_{s,d} = 110$ K) for Bi-2212 and Bi-2223, respectively (see Fig. 9(c)).

After turning the light on, an average temperature increase of 2.8 K is recorded at the sample position. Again, it can be checked for both phases that the expected increase in their normal in-plane resistivity is less than 2.5 % [16,29], so that, a part from the effect of photogenerated carriers, we can approximate the starting value of their resistivity with $\rho_{2,d} = \rho_2(T_{s,d} = 110$ K) and $\rho_{3,d} = \rho_3(T_{s,d} = 110$ K). On top of this, we have to add the effect of carriers created by the illumination for those sample regions that are reached by them. Accordingly with previous hypotheses, $N_{def}/2$ holes are injected in the Bi-2212 phase matrix for a length of $L - l_3$ and a width of $w_3$, providing the new resistivity value $\rho_{2,l}$ for this phase. Because of the presence of a non-vanishing electric field also in the Bi-2223 region, carriers are now injected in this region as well. This means that $N_{def}/2$ holes are injected in the Bi-2223 filament with a length of $l_3$ and a width of $w_3$, so that its new resistivity value is $\rho_{3,l}$. For simplicity, the same hypothesis of uniform distribution for the additional carriers will be used for the Bi-2223 filament, too. The final situation under these conditions is the one sketched in Fig. 9(d).

We can now proceed with the quantitative assessment of the model. Let's start with the analysis of $T_a = 100$ K in the dark (Fig. 9(a)). In this case the Bi-2223 filament is dissipationless and therefore shorts the corresponding portion of Bi-2212. Therefore, the measured electrical resistance is $R_d(T = 100K) = \rho_{2,d}(L - l_3)(Wt)^{-1}$. From the definition of effective sample resistivity $\rho_{eff} = R\,S/L$, it follows that:

$$\rho_{eff,d}(T = 100K) = \rho_{2,d}(L - l_3)/L \ . \quad (3)$$

Of course, this implies that $l_3$ can be univocally determined for each annealing step from the data shown in Fig. 2, if $\rho_{2,d}$ is known for $T_a = 100$ K. This is exactly the case thanks to the measurements already performed in single-phase Bi-2212 whiskers grown at the same temperature [16], whose value is reported in Table 2. It is worth stressing that, as a consequence, $l_3$ is not a free parameter of the model.

**Table 2**. Experimental values used in the implementation of the pseudo-1D electrical model. $L$, $W$, and $t$ represent the sample length, width and thickness, respectively; $\tau$ indicates the quasiparticle scattering time, while $\rho_{2,d}$ and $\rho_{3,d}$ the dark resistivity of Bi-2212 and Bi-2223, respectively.

| Physical quantity | Value | Source |
|---|---|---|
| $L$ | $147.9 \pm 1.2\ \mu m$ | Present experiment |
| $W$ | $12.38 \pm 0.07\ \mu m$ | Present experiment |
| $t$ | $2.87 \pm 0.09\ \mu m$ | Present experiment |
| $\tau$ | $2.3 \cdot 10^{-14}\ s$ | Ref. 19 |
| $\rho_{2,d} = \rho_2(T_{s,d} = 100\ K)$ | $3.08 \cdot 10^{-6}\ \Omega\,m$ | Ref. 16 |
| $\rho_{2,d} = \rho_2(T_{s,d} = 110\ K)$ | $3.37 \cdot 10^{-6}\ \Omega\,m$ | Ref. 16 |
| $\rho_{3,d} = \rho_3(T_{s,d} = 110\ K)$ | $6 \cdot 10^{-6}\ \Omega\,m$ | Ref. 29 |

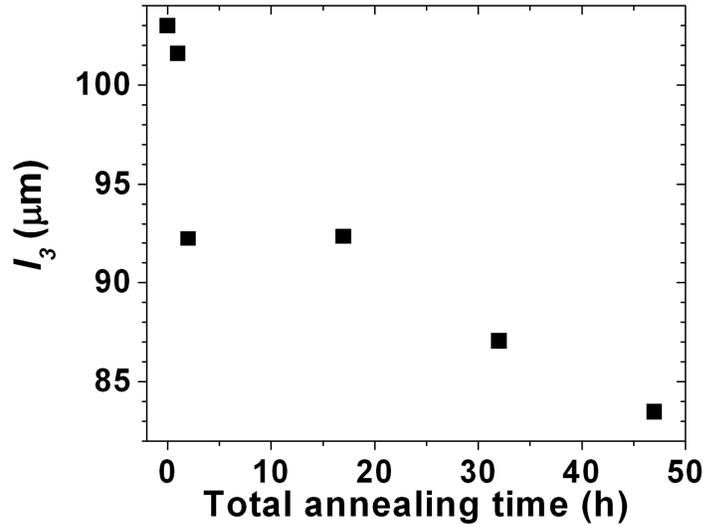

**Figure 10**. Evolution of the length $l_3$ of the equivalent Bi-2223 filament with the annealing time.

Figure 10 shows the results for the determination of $l_3$ as a function of the annealing time. A general trend corresponding to a decrease in the length of the Bi-2223 filament with increasing the annealing time can be clearly noticed. Such observation confirms the results of previous experiments in mixed-phase BSCCO whiskers, where it was shown that the Bi-2223 phase can decompose in favor of the Bi-2212 phase even at room temperature [13]. Moreover, it is well-known that this decomposition also results in mixed Ca and Cu oxides as by-products [30,31], which means that nanoinclusions of this kind are expected to originate in the regions where Bi-2223 converts into Bi-2212. These regions can certainly be considered as defects from the point of view of the conduction process in BSCCO and also represent good candidates as possible sites for electron trapping.

On the other hand, the situation at $T_a = 100$ K under illumination can be described as two resistors in series, the former represented by the short circuit corresponding to the Bi-2223 filament in the Meissner state, the latter arising from the parallel connection of the resistors corresponding to the regions with resistivity values $\rho_{2,d}$ and $\rho_{2,l}$, respectively (see Fig. 9(b)). Therefore, the overall electrical resistance of the sample is:

$$R_l(T=100K) = \left[ \left( \rho_{2,l} \frac{(L-l_3)}{w_3 t} \right)^{-1} + \left( \rho_{2,d} \frac{(L-l_3)}{(W-w_3)t} \right)^{-1} \right]^{-1}, \quad (4)$$

which implies for the effective resistivity in the light $\rho_{eff,l}$:

$$\rho_{eff,l}(T=100K) = R_l(T=100K)\frac{Wt}{L} = W\frac{L-l_3}{L}\frac{\rho_{2,l}\rho_{2,d}}{w_3\rho_{2,d}+(W-w_3)\rho_{2,l}} \quad . \quad (5)$$

An estimation of $\rho_{2,l}$ can now be done under the previous hypotheses. In a Fermi gas picture, the injection of $N_{def}/2$ holes induces an increase of the conductivity $\Delta\sigma_2 = \sigma_{2,l} - \sigma_{2,d} = \frac{N_{def}e^2\tau}{2Vm}$, where $e$ is the electron charge, $\tau$ is the quasiparticle elastic scattering time, $m$ the carrier effective mass and $V$ the volume into which the carriers are injected, that is in this case $V = (L-l_3)w_3 t$. The relationship between $\Delta\rho_2 = \rho_{2,l} - \rho_{2,d}$ and $\Delta\sigma_2$ corresponds to $\Delta\rho_2 = -\frac{\rho_{2,d}^2 \Delta\sigma_2}{1+\rho_{2,d}\Delta\sigma_2} \approx -\rho_{2,d}^2 \Delta\sigma_2$ for $\rho_{2,d}\Delta\sigma_2 \ll 1$, so that $\rho_{2,l}$ results as

$\rho_{2,l} = \rho_{2,d} - \rho_{2,d}^2 \frac{N_{def}e^2\tau}{2(L-l_3)w_3 tm}$ . By inserting this expression into Eq. (5) one obtains:

$$\rho_{eff,l}(T=100K) = W\frac{L-l_3}{L}\frac{\rho_{2,d} - \rho_{2,d}^2 \frac{N_{def}e^2\tau}{2(L-l_3)w_3 tm}}{W - (W-w_3)\rho_{2,d}\frac{N_{def}e^2\tau}{2(L-l_3)w_3 tm}} \quad . \quad (6)$$

The difference between Eq. (6) and Eq. (3) provides the analytical expression that should correspond to the curve for $T_a = 100$ K in Fig. 6(a):

$$\Delta\rho_{eff}(T=100K) = \rho_{eff,l} - \rho_{eff,d} = \rho_{2,d}\frac{L-l_3}{L}\left[\frac{W}{W-(W-w_3)A}(1-A) - 1\right] \quad , \quad (7)$$

where $A = \rho_{2,d}\frac{N_{def}e^2\tau}{2(L-l_3)w_3 tm}$.

Let us now analyze the situation with $T_a = 110$ K in the dark (see Fig. 9(c)). The difference with respect to $T_a = 100$ K is that the Bi-2223 filament is no longer in the superconducting state. Therefore a series connection of two non-vanishing resistors has to be considered for the equivalent circuit of the sample: the former is represented by the portion of Bi-2212 matrix with length $L–l_3$ and resistivity $\rho_{2,d} = \rho_2(T_{s,d} = 110$ K$)$, the latter results from the parallel combination of a resistor corresponding to the Bi-2223 phase with resistivity $\rho_{3,d} = \rho_3(T_{s,d} = 110$ K$)$ and of another resistor corresponding to the Bi-2212 matrix running along the filament. Therefore, according to the model the overall sample resistance is:

$$R_d(T=110K) = \rho_{2,d}\frac{(L-l_3)}{Wt} + \left[\left(\rho_{3,d}\frac{l_3}{w_3 t}\right)^{-1} + \left(\rho_{2,d}\frac{l_3}{(W-w_3)t}\right)^{-1}\right]^{-1}, \quad (8)$$

which corresponds to the following effective resistivity:

$$\rho_{eff,d}(T=110K) = R_d(T=110K)\frac{Wt}{L} = \rho_{2,d}\frac{(L-l_3)}{L} + \frac{\rho_{3,d}\rho_{2,d}}{w_3\rho_{2,d}+(W-w_3)\rho_{3,d}}\frac{Wl_3}{L}. \quad (9)$$

When light is turned on at $T_a = 110$ K the equivalent circuit becomes the parallel connection of a resistor representing the Bi-2212 matrix extending along the whole crystal length with resistivity $\rho_{2,d} \approx \rho_2(T_{s,d} = 110$ K) and of another resistor representing the series of the portion of the Bi-2212 matrix whose resistivity has decreased to $\rho_{2,l}$ because of the photogenerated holes and of the Bi-2223 filament with resistivity $\rho_{3,l}$ (see Fig. 9(d)). Therefore the sample resistance is:

$$R_l(T=110K) = \left[\left(\rho_{2,d}\frac{L}{(W-w_3)t}\right)^{-1} + \left(\rho_{2,l}\frac{(L-l_3)}{w_3 t} + \rho_{3,l}\frac{l_3}{w_3 t}\right)^{-1}\right]^{-1}, \quad (10)$$

which corresponds to the effective resistivity:

$$\rho_{eff,l}(T=110K) = R_l(T=110K)\frac{Wt}{L} = W\frac{\rho_{2,d}[\rho_{3,l}l_3 + \rho_{2,l}(L-l_3)]}{Lw_3\rho_{2,d}+(W-w_3)[\rho_{3,l}l_3+\rho_{2,l}(L-l_3)]}. \quad (11)$$

By using the usual hypothesis of uniform distribution of the additional carriers into the downstream portions of the sample, the resistivity under illumination of the Bi-2223 filament can be approximated with $\rho_{3,l} = \rho_{3,d} - \rho_{3,d}^2\frac{N_{def}e^2\tau}{2l_3 w_3 tm}$. Substitution of the proper relations for $\rho_{2,l}$ and $\rho_{3,l}$ into Eq. (11) provides the final expression:

$$\rho_{eff,l}(T=110K) = W\frac{\rho_{2,d}B}{Lw_3\rho_{2,d}+(W-w_3)B}, \quad (12)$$

with $B = \rho_{3,d}l_3 - \rho_{3,d}^2\frac{N_{def}e^2\tau}{2w_3 tm} + \rho_{2,d}(L-l_3) - \rho_{2,d}^2\frac{N_{def}e^2\tau}{2w_3 tm}$. The difference between Eq. (12) and Eq. (9) represents the result of the model that should correspond to the experimental curve for $T_a = 110$ K of Fig. 6(a).

Equations (7), (9) and (12) summarize the predictions of the model. We now discuss how the different parameters contained in these equations have to be considered in order to reproduce the experimental data of Fig. 6(a). The crystal sizes $L$, $W$ and $t$ have been measured by SEM and AFM and are reported in Table 2. The value of the hole effective mass $m$ has been assumed equal to the free electron mass for simplicity, while the elastic quasiparticle scattering time $\tau$ has been taken as the average value observed in paraconductivity experiments on whiskers grown with the same procedure [19]. This value is reported in Table 2 as well. Concerning $\rho_{3,d}(T = 110\ K)$, this quantity has been obtained from the measurements on Bi-2223 single crystals by Liang et al. [29], as listed in Table 2. Since $\rho_{2,d}$ and $l_3$ have already been discussed above, it is clear that the only free parameters of the model are $N_{def}$ and $w_3$, which can be determined by means of a comparison with the experimental behaviour of $\Delta\rho_{eff}(T = 100\ K)$ and $\Delta\rho_{eff}(T = 110\ K)$ as a function of the annealing time. In order to get better insight in the model features, we have treated $w_3$ as a fixed parameter by selecting *a priori* some value in a reasonable range and by keeping it constant during the whole annealing process. Since the most significant signal is $\Delta\rho_{eff}(T = 100\ K)$, the best fit value of $N_{def}$ has been determined for each data point of the curve, and then the corresponding $\Delta\rho_{eff}(T = 110\ K)$ has been calculated. The corresponding results are shown in Fig. 11 and Fig. 12, respectively.

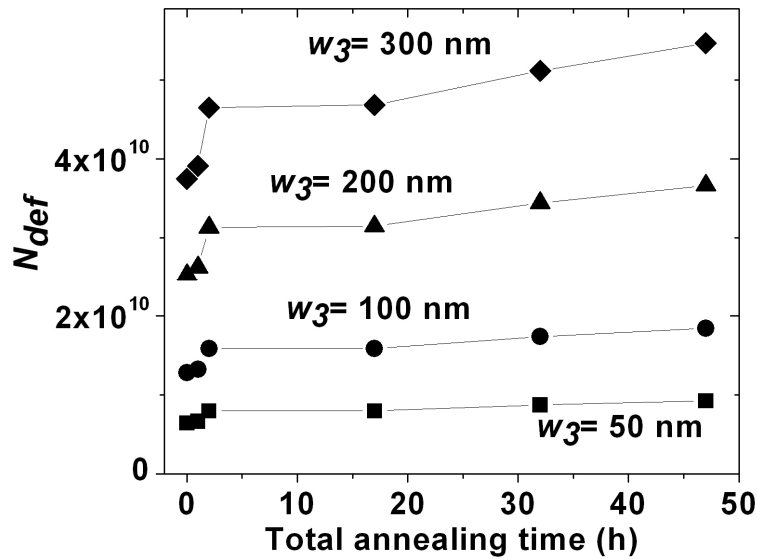

**Figure 11.** Evolution of best fit values of $N_{def}$ as a function of the annealing time for $w_3 = 50$ nm (squares), 100 nm (circles), 200 nm (triangles), and =300 nm (diamonds).

It is apparent that the values indicated for $N_{def}$ in Fig. 11 increase with increasing the annealing time, regardless of the value of $w_3$. This is in agreement with the fact that the decomposition of the Bi-2223 phase should also produce defective regions inside the freshly generated Bi-2212 domains. Moreover, the values obtained for $N_{def}$ compare favorably to the number of Bi-2212 crystal cells $N = 5.9 \cdot 10^{12}$ contained in the measured portion of the crystal, as determined by the ratio between its volume and the elementary cell volume $V_{cell} = 8.89 \cdot 10^{-28}$ m$^3$ measured via XRD, indicating that a fraction ranging form 0.1% to 0.9% of the cells should host a defect, depending on the specific values of $w_3$ and of the annealing time.

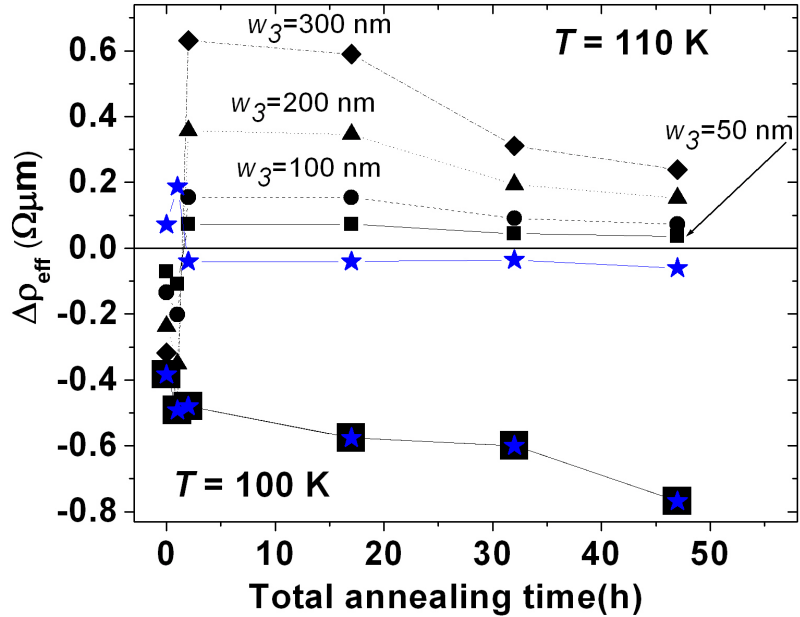

**Figure 12**. (Color online) Evolution of both experimental and calculated values for $\Delta\rho_{eff}(T = 100$ K$)$ and $\Delta\rho_{eff}(T = 110$ K$)$ as a function of the annealing time. Experimental values are represented by blue stars in both cases. Calculated values for $T = 100$ K are represented by big squares and overlap to experimental values for all of the $w_3$ values. Simulated values for $T = 110$ K are also shown for $w_3 = 50$ nm (small squares), 100 nm (circles), 200 nm (triangles) and 300 nm (diamonds).

Figure 12 summarizes the results of the model for $\Delta\rho_{eff}$. It is seen that, while the situation for $T = 100$ K is well reproduced by any value of $w_3$, for $T = 110$ K the best resemblance with the experimental data is obtained only for $w_3 = 50$-$100$ nm at most, i.e. with a width of the Bi-2223 filament that is only 0.4–0.8% of the crystal width. Such reasonable results confirm that the pseudo-1D model we have developed reflects the essential point of our experiment, considering also the several simplifications introduced. Some discrepancies between experimental and simulated data can still be observed, like for instance the sign of $\Delta\rho_{eff}$ at $T = 110$ K, and we think that this is probably due to fine details of the real distribution of the Bi-2223 filaments in the Bi-2212 matrix that have been neglected. In this respect, a likely oversimplification in our model could be represented by the assumption that the thickness of the Bi-2223 filaments always coincides with the one of the whole crystal.

## 5. Conclusions

We have carried out a complete set of photoconductivity experiments on BSCCO whiskers that have been aged by means of subsequent annealing steps. We have shown that in the case of BSCCO mixed-phase crystals a moderate enhancement in conductivity appears at some temperatures and that it evolves with the annealing of the samples. A simple model of the Kudinov type has been developed for the interpretation of photoconductivity data, which is based on the idea of the presence of filamentary inclusions of the Bi-2223 phase in the Bi-2212 matrix, as previously reported by other authors. This model accounts reasonably well for the photoconductivity experimental data and gives further insight into the process taking place during the ageing by showing that the length of the Bi-2223 inclusions shrinks and that the number of electronically active defects in the crystal increases with increasing the annealing time of the crystal. Such results imply that Ca- and Cu-rich nanometric defects resulting from Bi-2223 filament decomposition can act as effective photoelectron traps and suggest a way to modulate the photoconductivity response of Bi-2212 crystals by means of the controlled addition of nanoparticles of this kind. Finally, the non-persistent nature of photoconductivity in Bi-2212 indicates that this material could be more suitable than YBCO to fabricate non-hysteretic nanowire photon detectors.

**Acknowledgements**

We gratefully thank Marie Gabrielle Medici for allowing access to the laboratory and for help in data acquisition. We also acknowledge financial support by the project ORTO11RRT5 in the framework of Progetti di Ricerca di Ateneo – Compagnia di San Paolo – 2011.

**Appendix**

Eq. (1) results from considering the energy balance for the sample region under illumination. The geometry of such experimental situation is represented in Fig. 13. When considering the energy balance equation in steady state conditions $\dot{Q}_{in} = \dot{Q}_{out}$ for the cylindrical shell between $r$ and $r + dr$, one gets:

$$-K\left(\frac{dT}{dr}\right)_r 2\pi r\, l + P\, 2\pi r\, dr = -K\left(\frac{dT}{dr}\right)_{r+dr} 2\pi (r+dr)\, l \quad , \tag{13}$$

where $P$ is the power per unit surface provided by the light beam, $K$ is the copper thermal conductivity, $l$ is the copper head thickness, $r$ is the distance from the beam center. To the leading order, this equation can be rewritten as:

$$\frac{d^2T}{dr^2} + \frac{1}{r}\frac{dT}{dr} + \frac{P}{Kl} = 0 \quad ,$$

or, equivalently, as:

$$\frac{1}{r}\frac{d}{dr}\left(r\frac{dT}{dr}\right) + \frac{P}{Kl} = 0 \quad . \tag{14}$$

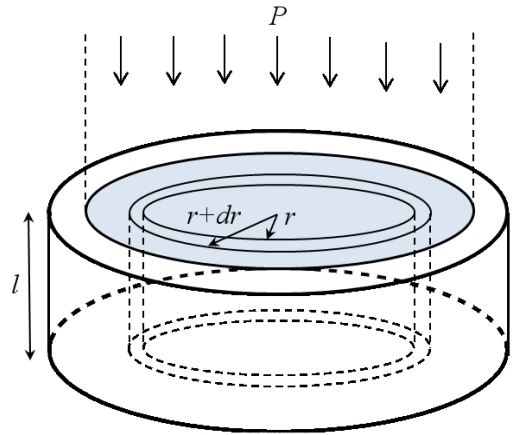

**Figure 13**. (Color online) Sketch of the geometrical model representing the cold head under illumination that has been used for determining the sample temperature correction.

This equation can be integrated by casting it into the form:

$$\frac{d}{dr}\left(r\frac{dT}{dr}\right) = -\frac{P}{Kl}r$$

which gives the general solution: $r\frac{dT}{dr} = -\frac{P}{2Kl}r^2 + B$, where $B$ is a constant whose value can be determined from the observation that, for symmetry reasons, the center of the beam $r=0$ must be a maximum of the temperature field. This implies $B = 0$, so that:

$$\frac{dT}{dr} = -\frac{P}{2Kl}r$$

and a new integration gives the final result:

$$T(r) = -\frac{P}{4Kl}r^2 + C \;.$$